\documentclass[epj]{webofc}
\usepackage[utf8]{inputenc}
\usepackage[varg]{txfonts}   % Web of Conferences font
\usepackage{booktabs}
\usepackage{xcolor}
\definecolor{darkred}{rgb}{0.4,0.0,0.0}
\definecolor{darkgreen}{rgb}{0.0,0.4,0.0}
\definecolor{darkblue}{rgb}{0.0,0.0,0.4}
\usepackage[bookmarks,linktocpage,colorlinks,
    linkcolor = darkred,
    urlcolor  = darkblue,
    citecolor = darkgreen]{hyperref}
\usepackage{subfigure}
\wocname{EPJ Web of Conferences}
\woctitle{Lattice2017}
\def\slash#1{\mbox{$\not \!\! #1$}}
\def\Dslash{{\slash {\cal D}}}

\def\lvec#1{\setbox0=\hbox{$#1$}
    \setbox1=\hbox{$\scriptstyle\leftarrow$}
    #1\kern-\wd0\smash{
    \raise\ht0\hbox{$\raise1pt\hbox{$\scriptstyle\leftarrow$}$}}
    \kern-\wd1\kern\wd0}
\def\rvec#1{\setbox0=\hbox{$#1$}
    \setbox1=\hbox{$\scriptstyle\rightarrow$}
    #1\kern-\wd0\smash{
    \raise\ht0\hbox{$\raise1pt\hbox{$\scriptstyle\rightarrow$}$}}
    \kern-\wd1\kern\wd0}
\def\diracstar#1#2{
    \setbox0=\hbox{$\gamma$}\setbox1=\hbox{$\gamma_{#1}$}
    \gamma_{#1}\kern-\wd1\kern\wd0
    \smash{\raise4.5pt\hbox{$\scriptstyle#2$}}}
\def\tr{\,\hbox{tr}\,}

\newcommand{\beq}{\begin{equation}}
\newcommand{\eeq}{\end{equation}}
\newcommand{\beqn}{\begin{eqnarray}}
\newcommand{\eeqn}{\end{eqnarray}}
\newcommand{\nn}{\nonumber}

\begin{document}
%%%%%%%%%%%%%%%%%%%%%%%%%%%%%%%%%%%%%%%%%%%%%%%%%%%%%%%%%%%%%%%%%%%%%%%%%%%%%
%
\selectlanguage{english}
%----------------------------------------------------------------------------
\title{%
Testing a non-perturbative mechanism for elementary fermion mass generation: lattice setup}
%----------------------------------------------------------------------------
\author{%
\firstname{Stefano} \lastname{Capitani}\inst{1} \and
\firstname{Giulia Maria} \lastname{de Divitiis}\inst{2} \and
\firstname{Petros}  \lastname{Dimopoulos}\inst{2,3}\and
\firstname{Roberto} \lastname{Frezzotti}\inst{2} \and
\firstname{Marco} \lastname{Garofalo}\inst{4} \fnsep\thanks{Speaker, \email{s1459858@sms.ed.ac.uk}}\and
\firstname{Bastian} \lastname{Knippschild}\inst{5} \and
\firstname{Bartosz} \lastname{Kostrzewa}\inst{5} \and
\firstname{Ferenc} \lastname{Pittler}\inst{5} \and
\firstname{Giancarlo} \lastname{Rossi}\inst{2,3} \and
\firstname{Carsten} \lastname{Urbach}\inst{5} 
% etc.
}
%----------------------------------------------------------------------------
\institute{%
Johann Wolfgang Goethe-Universit\"at Frankfurt am Main, Institut f\"ur Theoretische Physik, Max-von-Laue-Stra\ss e 1, D-60438 Frankfurt am Main,
Germany \and
 Dipartimento di Fisica, Universit\`a di  Roma  ``{\it Tor Vergata}'' and INFN, Sezione di Roma 2,     Via della Ricerca Scientifica - 00133 Rome, Italy\and
 Centro Fermi - Museo Storico della Fisica e Centro Studi e Ricerche Enrico Fermi, Compendio del Viminale, Piazza del Viminiale 1, I-00184, Rome, Italy
 \and
Higgs Centre for Theoretical Physics, School of Physics and Astronomy,    The University of Edinburgh, Edinburgh EH9 3JZ, Scotland, UK  \and
   Helmholtz Institut f\"ur Strahlen-und Kernphysik (Theorie), Nussallee 14-16 Bethe Center for Theoretical Physics, Nussallee 12 Universit\"at Bonn, D-53115 Bonn, Germany
 }
%----------------------------------------------------------------------------
\abstract{%
  In this contribution we lay down a lattice setup that allows for the non-perturbative study of a
  field theoretical model where a SU(2) fermion doublet, subjected to non-Abelian gauge interactions,
  is also coupled to a complex scalar field doublet via a Yukawa and an “irrelevant” Wilson-like 
  term. Using naive fermions in quenched approximation and based on the renormalized Ward identities
  induced by purely fermionic chiral transformations, lattice observables are discussed that enable:
  a) in the Wigner phase, the determinations of the critical Yukawa coupling value where the purely
  fermionic chiral transformation become a symmetry up to lattice artifacts; b) in the Nambu-Goldstone
  phase of the resulting critical theory, a stringent test of the actual generation of a fermion mass
  term of non-perturbative origin. A soft twisted fermion mass term is introduced to circumvent the
  problem of exceptional configurations, and observables are then calculated in
  the limit of vanishing twisted mass.
}
%----------------------------------------------------------------------------
\maketitle
%----------------------------------------------------------------------------

\section{Introduction}\label{sec:intro}
In~\cite{Frezzotti:2014wja} a new non-perturbative (NP) mechanism for elementary
particle mass generation was conjectured. Existence and main properties of this phenomenon can be tested in the toy model described by the Lagrangian
\beqn
&&{\cal L}_{\rm{toy}}(\Psi,A,\Phi)= {\cal L}_{kin}(\Psi,A,\Phi)+{\cal V}(\Phi)
+{\cal L}_{Wil}(\Psi,A,\Phi) + {\cal L}_{Yuk}(\Psi,\Phi) \, ,\label{SULL} \\
&&\quad{\cal L}_{kin}(\Psi,A,\Phi)= \frac{1}{4}(F\cdot F)+\bar \Psi_L\Dslash \Psi_L+\bar \Psi_R\Dslash \,\Psi_R+\frac{1}{2}{\tr}\big{[}\partial_\mu\Phi^\dagger\partial_\mu\Phi\big{]}\label{LKIN}\\
&&\quad{\cal V}(\Phi)= \frac{\mu_0^2}{2}{\tr}\big{[}\Phi^\dagger\Phi\big{]}+\frac{\lambda_0}{4}\big{(}{\tr}\big{[}\Phi^\dagger\Phi\big{]}\big{)}^2\label{LPHI}\\
&&\quad{\cal L}_{Wil}(\Psi,A,\Phi)= \frac{b^2}{2}\rho\,\big{(}\bar \Psi_L{\overleftarrow{\cal D}}_\mu\Phi {\cal D}_\mu \Psi_R+\bar \Psi_R \overleftarrow{\cal D}_\mu \Phi^\dagger {\cal D}_\mu \Psi_L\big{)}
\label{LWIL} \\
&&\quad{\cal L}_{Yuk}(\Psi,\Phi)=\
  \eta\,\big{(} \bar \Psi_L\Phi \Psi_R+\bar \Psi_R \Phi^\dagger \Psi_L\big{)}
\label{LYUK} \, ,
\eeqn
where $b^{-1}=\Lambda_{UV}$ is the UV-cutoff.
The Lagrangian~(\ref{SULL}) describes a
SU(2) fermion doublet subjected to non-Abelian gauge interaction and coupled to a complex scalar field via Wilson-like
(eq.~(\ref{LWIL})) and Yukawa (eq.~(\ref{LYUK})) terms. For short we use a compact SU(2)-like notation where 
$\Psi_L=(u_L\,\,d_L)^T$ and $\Psi_R=(u_R\,\,d_R)^T$ are fermion iso-doublets and $\Phi$ is a $2\times2$
matrix with $\Phi=(\phi,-i\tau^2 \phi^*)$ and $\phi$ an iso-doublet of complex scalar fields.
The term ${\cal V}(\Phi)$ in eq.~(\ref{LPHI}) is the standard quartic scalar potential where the 
(bare) parameters $\lambda_0$ and $\mu_0^2$ control the self-interaction and the mass of the scalar
field. In the equations above we have introduced the covariant derivatives
\beq
{\cal D}_\mu=\partial_\mu -ig_s \lambda^a A_\mu^a \, , \qquad
\overleftarrow{\cal D}_\mu =\overleftarrow{\partial}_\mu +ig_s \lambda^a A_\mu^a \, ,\label{COVG}
\eeq
where $A_\mu^a$ is the gluon field ($a=1,2,\dots, N_c^2-1$) with field strengt $F_{\mu\nu}^{a}$. . The model~(\ref{SULL}) is power-counting renormalizable (as LQCD is) with counter-terms constrained
by the exact symmetries of the Lagrangian. Besides Lorentz, gauge and $C$, $P$, $T$, $CPF_2$ symmetries (see Appendix B of~\cite{Frezzotti:2014wja}),
${\cal L}_{\rm toy}$ is invariant under the following (global) transformations $\chi_L$ and $\chi_R$
\beqn
&&\bullet\,\chi_{L}:\quad \tilde\chi_{L}\otimes (\Phi\to\Omega_L\Phi) 
\quad\quad\bullet\,\chi_{R}:\quad \tilde\chi_{R}\otimes (\Phi\to\Phi\Omega_R^\dagger) 
\label{CHIL}\\
%\hspace{-2.cm}{\mbox{where}}\nn\\
\hspace{-2.cm}&&\tilde\chi_{L/R} : \left \{\begin{array}{l}     
\Psi_{L/R}\rightarrow\Omega_{L/R} \Psi_{L/R}  \\
\hspace{4cm}\Omega_{L/R}\in {\mbox{SU}}(2)_{L/R}\\
\bar \Psi_{L/R}\rightarrow \bar \Psi_{L/R}\Omega_{L/R}^\dagger \\ 
\end{array}\right. , \label{GTWT}
\eeqn
which forbid power divergent fermion mass terms.
The $d=4$ Yukawa term ${\cal L}_{Yuk}$ and the Wilson-like $d=6$ operator  ${\cal L}_{Wil}$, which for dimensional reasons enters in the Lagrangian multiplied by $b^2$,
break explicitly chiral transformations $\tilde\chi_L$  and $\tilde\chi_R$.
To study possible enhancement of $\tilde\chi_L$
symmetry (by parity the same will hold also for $\tilde\chi_R$) we consider the bare
Schwinger Dyson Equation (SDE)
\beqn
\partial_\mu \langle \tilde J^{L\, i}_\mu(x) \,\hat{\cal O}(0)\rangle = \langle \tilde\Delta_{L}^i \hat{\cal O}(0)\rangle\delta(x) -
\eta \,\langle \big{(} \bar \Psi_L\frac{\tau^i}{2}\Phi \Psi_R-\bar \Psi_R\Phi^\dagger\frac{\tau^i}{2}\Psi_L \big{)}(x)\,\hat{\cal O}(0)\rangle \!+\nn\\
\phantom{\partial_\mu J^{L\, i}_\mu}-\frac{b^2}{2}\rho\,\langle\Big{(} \bar \Psi_L\overleftarrow {\cal D}_\mu\frac{\tau^i}{2}\Phi{\cal D}_\mu \Psi_R-\bar \Psi_R\overleftarrow {\cal D}_\mu \Phi^\dagger\frac{\tau^i}{2}{\cal D}_\mu \Psi_L\Big{)}(x)\,\hat{\cal O}(0)\rangle\, ,\label{CTLTI} 
\eeqn
where $\tilde\Delta_{L}^i\hat{\cal O}(0)$ is the variation of $\hat{\cal O}(0)$ under $\tilde\chi_L$ and the associated non-conserved currents  are %to the 
\beq
\tilde J_\mu^{L\,i}= \bar \Psi_L\gamma_\mu\frac{\tau^i}{2}\Psi_L -\frac{b^2}{2}\rho\Big{(}\bar \Psi_L\frac{\tau^i}{2}\Phi {\cal D}_\mu \Psi_R - \bar \Psi_R\overleftarrow {\cal D}_\mu\Phi^\dagger\frac{\tau^i}{2} \Psi_L\Big{)}\, .
\label{JCLT}
\eeq
Under renormalization the $d = 6$ operator 
$ \; O_{6}^{L\, i} = \frac{1}{2} \rho
\Big{[}\bar \Psi_L\overleftarrow {\cal D}_\mu\frac{\tau^i}{2}\Phi{\cal D}_\mu \Psi_R
-{\mbox {h.c.}}\Big{]} \; $ 
mixes with two $d = 4$ operators, plus a set of six-dimensional ones that we globally denote by $[O_{6}^{L\, i}]_{sub}$~\footnote{We do not need to resolve the mixing among the different $d=6$ operators, as they only yield negligible O($b^2$) effects. To simplify the mixing pattern (\ref{O6L-MIX}) we used $\partial_\mu J^{L,i}_{\mu}=0$, where $J_\mu^{L,i}$ is the Noether current associated with the exact symmetry $\chi_L$ (\ref{CHIL}).}, viz.
\beq
 O_{6}^{L\, i} = 
\Big{[} O_{6}^{L\, i}  \Big{]}_{sub} +
\frac{Z_{\partial\tilde{J}}-1}{b^{2}}\partial_\mu\tilde J^{L\, i}_\mu
-\frac{\bar\eta}{b^{2}}\Big{[}\bar \Psi_L\frac{\tau^i}{2}\Phi \Psi_R
-{\mbox {h.c.}}\Big{]} + \ldots
\label{O6L-MIX}
\eeq
where $Z_{\partial\tilde{J}}$ and $\bar\eta$ are functions of the dimensionless bare parameters entering~(\ref{SULL}) and hence depend on the subtracted scalar squared mass $\mu_{sub}^2=\mu_0^2-b^2\tau$  through the combination $b^2\mu_{sub}^2$ that is a negligible $O(b^2 )$ quantity \cite{Frezzotti:2014wja}. Thus we write 
$Z_{\partial\tilde{J}}=Z_{\partial\tilde{J}}(\eta;g^2_s,\rho,\lambda_0)$ and $\bar\eta=\bar\eta(\eta;g^2_s,\rho,\lambda_0)$. Ellipses in the r.h.s.\ of eq.~(\ref{O6L-MIX}) denote possible NP contributions to operator mixing, the possible occurrence of which is a key point that will be discussed below. Plugging  (\ref{O6L-MIX}) into (\ref{JCLT}) we get 
\beq
 \partial_\mu \langle Z_{\partial\tilde{J}}\tilde J^{L,i}_\mu(x) \,\hat {\cal O}(0)\rangle\! = \!\langle \tilde\Delta_{L}^i \hat {\cal O}(0)\rangle\delta(x) -
({\eta- \overline\eta}) \,\langle {O_{Yuk}^{L,i}}(x)\,\hat {\cal O}(0)\rangle +{\ldots}+{\mbox{O}(b^2)}\nn. \label{ren_WTI}
\eeq
We define $\eta_{cr}(g^2_s,\rho,\lambda_0)-\bar\eta(\eta_{cr}; g^2_s,\rho,\lambda_0)=0$.
Setting $\eta=\eta_{cr}(g^2_s,\rho,\lambda_0)$ the SDE takes the form of a WTI
\beq
 \partial_\mu \langle Z_{\partial\tilde{J}}\tilde J^{L,i}_\mu(x) \,\hat {\cal O}(0)\rangle\! = \!\langle \tilde\Delta_{L}^i \hat {\cal O}(0)\rangle\delta(x) 
 +{\ldots}+{\mbox{O}(b^2)}\label{SYMCH} \, ,
\eeq
implying restoration of the fermionic $\tilde\chi_L \otimes \tilde\chi_R$ symmetries up to O($b^2$)  UV cutoff effects.

\subsection{Mass generation mechanism in the critical model (Nambu-Goldstone phase)} %!!!
The physics of the model~(\ref{SULL}) at the critical value $\eta_{cr}$ crucially depends on whether the parameter $\mu^2_0$ is such that ${\cal V}(\Phi)$ has a unique minimum (Wigner phase of the $\chi_L$ symmetry, $\mu_{sub}^2 > 0$) or whether ${\cal V}(\Phi)$ develops the typical ``mexican hat'' shape (Nambu-Goldstone phase $\mu_{sub}^2 < 0$). 
Here $\mu_{sub}^2=\mu_0^2-\mu_{cr}^2$, with $\mu_{cr}^2$ being the phase transition point.
%%\begin{itemize}
%%\item 
In the Wigner phase no NP terms (i.e.\ ellipses) are expected to occur in the mixing pattern of eq.~(\ref{O6L-MIX}) and the transformations $\tilde \chi_L$ leads to eq.~(\ref{SYMCH}) without the ellipses \citep{Frezzotti:2014wja}.

%\beqn\hspace{-1.4cm}&&\partial_\mu \langle Z_{\tilde J}\tilde J^{L\, i}_\mu(x) \,\hat O(0)\rangle\Big{|}_{\eta_{cr}} = \langle \tilde\Delta_{L}^i\hat O(0)\rangle\Big{|}_{\eta_{cr} \delta(x)+{\mbox O}(b^2)\, .\label{CTLTI-RCR}\eeqn

\begin{figure}[thb] % no figure before 1st section
  \centering
  \includegraphics[width=\textwidth,clip]{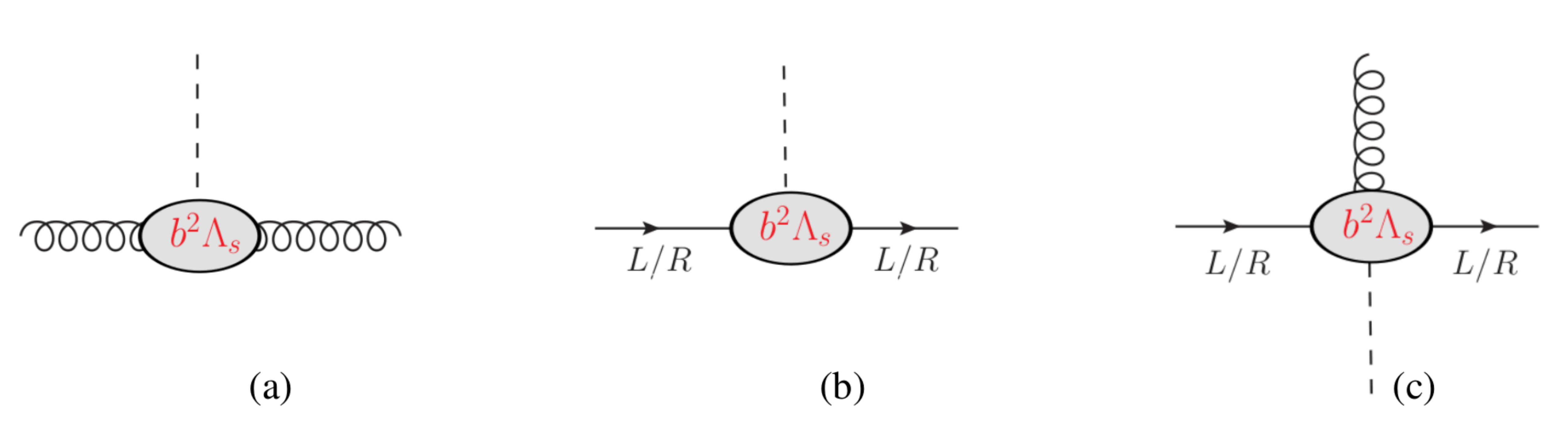}
  \caption{Some of the NP O($b^2\Lambda_s\alpha_s^2$) effective vertices that are conjectured to arise~\cite{Frezzotti:2014wja} in the Nambu-Goldstone phase of the model.}
  \label{fig:vertices}% Give a unique label
\end{figure}

\begin{figure}[thb] % no figure before 1st section
  \centering
  \includegraphics[width=\textwidth,clip]{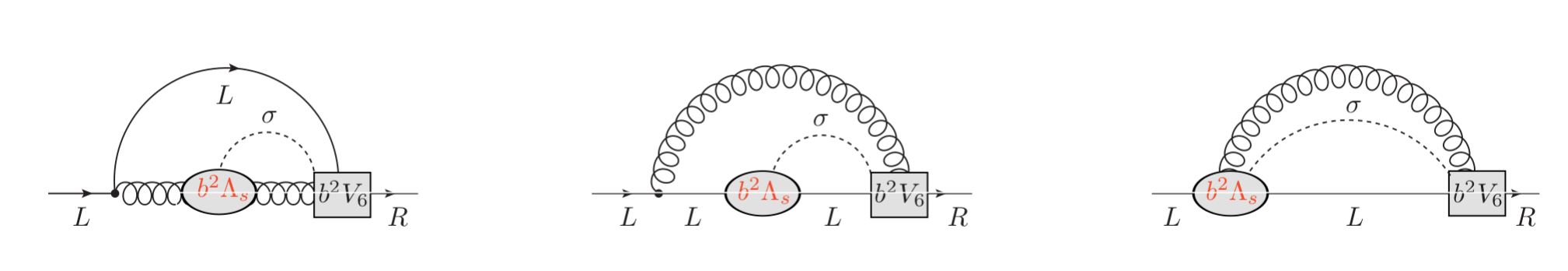}
 \caption{Typical lowest order self-energy "diagrams" giving rise to dynamically generated
quark mass terms. The grey box represents the insertion of the Wilson-like vertex stemming from ${\cal L}_{Wil}$. The dotted line represents the propagation of a scalar particle. The $b^{-4}$ loop divergency is cancelled by the two vertices O$(b^2)$ giving rise to a finite result.}
\label{fig:self_energy}
\end{figure}

%%\item 
In the Nambu-Goldstone phase a non-perturbative term is expected/conjectured  
\cite{Frezzotti:2014wja} to appear in the mixing pattern of eqs.~(\ref{O6L-MIX}) leading to a WTI of the form
\beq
  \partial_\mu \langle Z_{\partial\tilde J}\tilde J^{L,i}_\mu(x) \,\hat {\cal O}(0)\rangle_{\eta_{cr}} = \langle \tilde\Delta_{L}^i \hat {\cal O}(0)\rangle_{\eta_{cr}}\delta(x)+\langle {C_1\Lambda_s[ \overline \Psi_L \frac{\tau^i}{2}{\cal U} \Psi_R+\mbox{h.c.}]} \hat {\cal O}(0)\rangle +{\mbox O}(b^2)
  \label{partialJ_continuum_NG}
\eeq
where
\beq
{\cal U} = \frac{\Phi}{\sqrt{\Phi^\dagger \Phi}}=\frac{v+\sigma+i\vec{\tau}\vec{\pi}}{\sqrt{(v+\sigma)^2+\vec{\pi}\vec{\pi}}}\, .\label{U}
\eeq 
${\cal U}$ is a dimensionless non-analytic function of $\Phi$ that has the same transformation properties as the latter under $\chi_L \times \chi_R$ and is well defined only if $\langle\Phi\rangle=v\neq 0$.
Occurrence of the $c_1 \Lambda_s$ term in the (\ref{partialJ_continuum_NG}) implies the presence of $c_1 \Lambda_s \bar \Psi \Psi$ term
in $\Gamma^{loc}_{NG}$, the local effective action in the NG phase.
This term describe NP breaking of $\tilde \chi_L\otimes \tilde \chi_R$ and in particular gives fermions a mass $c_1\Lambda_s$. 
It does not stem from the Yukawa term and, interestingly, can give a natural  (in the sense of 't~Hooft~\cite{tHooft:1979rat})  understanding of the fermion mass hierarchy problem (see discussion in \cite{Frezzotti:2014wja}).
An idea of how the mechanism works can be obtained from a perturbative expansion where Feynman diagrams are evaluated with the Lagrangian (\ref{SULL}) augmented 
by few extra terms representing the expected $O(b^2)$ NP effective vertices \cite{Frezzotti:2014wja}, as those shown in fig.~\ref{fig:vertices}.
These vertices can be inserted together with $O(b^2)$ vertices coming from the term (\ref{LWIL}) in diagrams like the ones depicted in fig.~\ref{fig:self_energy}, giving rise to finite self-energy contributions.\\
It is worth noticing that if the mechanism we have conjectured really exists it will generate a NP mass term for the fermions even in the quenched approximation where the vertices (b) and (c) of fig.~\ref{fig:vertices}, and thus the two rightmost diagrams of fig.~\ref{fig:self_energy}, are still present.
\vspace{-0.2cm}

\section{Lattice quenched study of ${\cal L}_{toy}$: regularization and renormalization }

Numerical simulations of lattice models with gauge, fermions and scalars are not common and technically 
% !!!! challenging\footnote{To our knowledge this is the first numerical study of a model with strongly interacting gauge fields, fermions and scalars.}. 
challenging\footnote{To our knowledge what we presented here is the first numerical study of a model with fermions, scalars and non-Abelian gauge fields in the strong interaction regime.}. 
In this first numerical study of the model~(\ref{SULL}) we can limit ourselves to a 
{\em quenched-fermion} simulation of the lattice model specified below. In fact in quenched approximation the gauge and the scalar fields can be updated independently
of each other. The lattice regularized action
\footnote{For a presentation of preliminary numerical results see \cite{Petros}}
 we consider reads
\begin{eqnarray}
&&{ { S}_{{ lat}}=b^4\sum_x\Big{\{}{\cal L}_{kin}^{YM}[U]+{\cal L}_{kin}^{sca}(\Phi)+{\cal V}(\Phi)}{+ \overline  \Psi  { D_{lat}[U,\Phi]}  \Psi\Big{\}}}  
\label{L_lat}\\ 
 &&{\cal L}_{kin}^{YM}[U]\,\,{\mbox{: SU($3$) plaquette action}}  
\\ 
 &&{\cal L}_{kin}^{sca}(\Phi)+{\cal V}(\Phi)=\frac{1}{2}{\tr}[\Phi^\dagger(-\partial_\mu^*\partial_\mu)\Phi]+\frac{\mu_0^2}{2}{\tr}\big{[}\Phi^\dagger\Phi\big{]}+\frac{\lambda_0}{4}\big{(}{\tr}\big{[}\Phi^\dagger\Phi\big{]}\big{)}^2 \,,
\end{eqnarray}
where $\Phi = \varphi_01\!\!1+i \varphi_j\tau^j$ is  a matrix-valued field and 
${\tr}[\Phi^\dagger\Phi]=\varphi_0^2+\varphi_1^2+\varphi_2^2+\varphi_3^2$
\begin{eqnarray}
 (D_{lat}[U,\Phi]&&\!\! \Psi)(x)=
\gamma_\mu \widetilde\nabla_\mu  \Psi(x) + \eta F(x)  \Psi(x)
-b^2\rho \frac{1}{2}F(x)
\widetilde\nabla_\mu \widetilde\nabla_\mu  \Psi(x)
\\ 
&& - b^2\rho \frac{1}{4} \Big{[} (\partial_\mu F)(x) U_\mu(x) \widetilde\nabla_\mu  \Psi(x+\hat\mu) + (\partial_\mu^* F)(x) U_\mu^\dagger(x-\hat\mu) \widetilde\nabla_\mu  \Psi(x-\hat\mu) \Big{]} \,,
\end{eqnarray}
 with $\; F(x) \equiv [\varphi_0 1\!\!1
+i\gamma_5\tau^j\varphi_j](x) $, the fermionic SU(2) doublet $\Psi^T=( u,d)$ and the lattice derivatives defined as
\begin{eqnarray}
&&\nabla_\mu f(x)\equiv \frac{1}{b}(U_\mu(x)f(x+\hat\mu)-f(x))\quad\quad\nabla_\mu^* f(x)\equiv \frac{1}{b}(f(x)-U_\mu^\dagger(x-\hat\mu) f(x-\hat\mu))\label{forward_back_derivative}\\
&&\widetilde\nabla_\mu f(x)\equiv \frac{1}{2}(\nabla_\mu+\nabla_\mu^*)F(x).
\end{eqnarray}

The Wilson-like term does not remove the doublers because it involves the scalar field  $\Phi$ and it has dimension six. This makes no harm in this quenched study aimed at testing whether the mass generation mechanism occurs at all.
The analysis done in \cite{KlubergStern:1983dg},
\citep{Patel:1992vu} and \citep{Luo:1996vt} for staggered fermions can be used to analyze the fermions in the Lagrangian (\ref{L_lat}):
first we rewrite the action in terms of the   field
$\chi(x)={\cal A}_x^{-1}\Psi(x)$ with  ${\cal A}_x=\gamma_1^{x_1}\gamma_2^{x_2}\gamma_3^{x_3}\gamma_4^{x_4}$, then
we perform a second change of variables  
\begin{equation}
q^B_{\alpha,a}(y)=\frac{1}{8}\sum_\xi \overline{U}(2y,2y+\xi) [\Gamma_\xi]_{\alpha,a}(1-b\sum_\mu\xi_\mu\widetilde\nabla_{\mu})\chi^B(2y+\xi), 
\end{equation}
where $y$ runs over  the coarse lattice  $ x_\mu=2y_\mu+\xi_\mu$, $\xi_\mu=0,1$ and $\overline{U}(2y,2y+\xi)$ is the average of link products along the shortest paths from $y$ to
$y + \xi$. With these changes of variables the action becomes  
 \begin{equation}\label{S_latt[q]}
 S_{lat}^{fer}=\sum_{y,B}\bar q^B(y)\left\{\sum_\mu (\gamma_\mu \otimes 1\!\!1)D_\mu +(\eta-\bar \eta){\cal F}(y) \right\} q^B(y)+O(b^2) \\ 
\end{equation}
where 
$ {\cal F}(y)=\varphi_0(2y)(1\!\!1 \otimes 1\!\!1)+s_Bi\tau^i\varphi_i(2y)(\gamma_5 \otimes t_5)$, $s_B=\pm 1$ and $t_\mu=\gamma_\mu^*$ are the taste matrices. The action (\ref{S_latt[q]}) is diagonal in 
taste and replicas $B=1,2,3,4$ indices
 up to $O(b^2)$; it describes $32$ fermions species namely $4$ replicas of $4$ tastes of the SU(2) doublet $q^T=(u,d)$.
\\
The quark bilinears in the $\Psi$ basis have well defined quantum numbers in the classical continuum limit once expressed in the $q^B$ basis.
For example the point-split vector current 
\begin{equation}
\widetilde J_\mu^{V^i}(x)=\overline\Psi(x-\hat\mu)\gamma_\mu\frac{\tau^i}{2}U_\mu(x-\hat\mu)\Psi(x)+\overline\Psi(x)\gamma_\mu\frac{\tau^i}{2}U_\mu^\dagger(x-\hat\mu)\Psi(x-\hat\mu)\,,\quad x_\mu=2y_\mu+\xi_\mu\,,
\end{equation}
once summed over the hypercube coordinate $\xi$ and expressed in the $q^B$ basis becomes
\begin{equation}\label{J_V}
\sum_\xi \widetilde J_\mu^{V^i}(2y+\xi) =\sum_{B=1}^4 \overline q^{B}(y)(\gamma_\mu\otimes 1\!\!1 )\frac{\tau^i}{2} q^B(y)+O(b^2).
\end{equation}
%Further we extend the Lagrangian (\ref{L_lat}) to two generations of fermions $\overline \Psi_\ell  D_{latt} \Psi_\ell+\overline \Psi_h  D_{latt} \Psi_h$;
%in this way disconnected diagrams are not present in the correlators between   operator off diagonal in generation space like e.g.
%\begin{equation}
%\sum_{\xi_1,\xi_2}\langle \overline\Psi_\ell(2y_1+\xi_1)\Gamma \tau \Psi_h(2y_1+%\xi_1)\overline\Psi_h(2y_2+\xi_2)\Gamma \tau \Psi_\ell(2y_2+\xi_2)\rangle \quad\quad \ell=(u,d)\quad h=(c,s)\,. 
%\end{equation}
One can prove that loop effects do not generate $d\leq 4$ operators besides
$ F_{\mu\nu}F_{\mu\nu}$, $ \partial_\mu\Phi^\dagger\partial_\mu\Phi $,  
$q^B(\gamma_\mu\otimes 1\!\!1)\tilde\nabla q^B$, $\Phi^\dagger \Phi$, $  (\Phi^\dagger \Phi)^2$ and $ 
\eta \overline q^B(y){\cal F}^B(y) q^B(y)$ which are all present in the action (\ref{L_lat}). A way of seeing this is based on "spectrum doubling symmetry" \cite{montvay_munster_1994}  
\begin{eqnarray}
\Psi(x) \to \Psi'(x)=e^{-ix\cdot\pi_H}M_H\Psi(x) \quad\quad \overline\Psi(x)\to \overline\Psi'(x)=\overline\Psi(x)M_H^\dagger e^{ix\cdot\pi_H}
\end{eqnarray}
where $H$ is an ordered set of four-vectors indices $H\equiv\{\mu_1,...,\mu_h\},\,(\mu_1<\mu_2<...<\mu_h)$. For $0\leq h \leq 4$ there are 16 four-vectors $\pi_H$ with $\pi_{H,\mu}=\pi$ if $\mu\in H$ or $\pi_{H,\mu}=0$ otherwise and 16 matrices
$M_H\equiv (i\gamma_5\gamma_{\mu_1})...(i\gamma_5\gamma_{\mu_h})$. 
This is an exact symmetry of $S_{lat}$, thus also of the effective action $\Gamma_{lat}[U,\Phi,\Psi]$. 
Now in order to respect the spectrum doubling symmetry $\Gamma_{lat}$ can only have 
 terms with symmetric covariant derivatives $\widetilde \nabla_\mu$ acting on $\Psi$.
Close to the continuum limit among the local terms of $\Gamma_{lat}$ only 
the fermion kinetic term $\bar\Psi \widetilde \nabla \Psi$ and Yukawa term $\eta\bar\Psi \Phi \Psi$ are relevant.

As a consequence we find that $\eta_{cr}$, the critical value of $\eta$,  
%at which $\tilde\chi_L \otimes \tilde\chi_R$ symmetry is restored up to O($b^2$)
is well defined (even in the presence of fermion doubling), unique and independent
of the subtracted scalar squared mass $\mu_{sub}^2$ (thus equal for 
the Wigner phase and the Nambu-Goldstone phase).

Since we are doing a quenched study of the model (\ref{L_lat}) exceptional configurations of the gauges fileds  and the scalars
with small eigenvalues of $D_{lat}$ can occur in the Monte Carlo sampling
 leading to small eigenvalues of $D_{lat}$. In order to get control over exceptional configurations we add a twisted mass term in the action
 \begin{equation}
 S_{lat}^{toy+tm}=S_{lat}+i\mu b^4\sum_x\overline \Psi \gamma_5\tau_3\Psi
 \end{equation}
  at the price of introducing a soft (hence harmless) breaking of $\chi_{L,R}$ (and $\widetilde \chi_{L,R}$ when restored). 

\section{Strategy of numerical study}
\label{Simulations strategy}
To study whether the NP mechanism occurs
we consider the renormlize axial $\widetilde\chi$ SDE (see eq.~\ref{partialJ_continuum_NG} )
\begin{equation}
 Z_{\partial\tilde A}\partial_\mu \widetilde  J_\mu^{A^\pm}=2(\eta -\eta_{cr}) \widetilde D^{P\pm}+ \delta_{ph,NG} C_1\Lambda_s{\cal P}^\pm+O(b^2)
 \label{axial_SDE}
\end{equation}
with $\delta_{ph,NG}=0,1$ for the NG and Wigner phase respectively, the current
\begin{eqnarray}
&&\widetilde J_\mu^{A^i}(x)=\overline\Psi(x-\hat\mu)\gamma_\mu\gamma_5\frac{\tau^i}{2}U_\mu(x-\hat\mu)\Psi(x)+\overline\Psi(x)\gamma_\mu\gamma_5\frac{\tau^i}{2}U_\mu^\dagger(x-\hat\mu)\Psi(x-\hat\mu) 
\end{eqnarray}
and the densities
\begin{equation}
\widetilde D^{P\pm}=\overline\Psi_L\left\{ \Phi,\frac{\tau^\pm}{2} \right\} \Psi_R-
\overline\Psi_R\left\{ \frac{\tau^\pm}{2},\Phi^\dagger \right\} \Psi_L\,, \quad\quad
 {\cal P}^\pm=\overline\Psi_L\left\{ {\cal U},\frac{\tau^\pm}{2} \right\} \Psi_R-
 \overline\Psi_R\left\{ \frac{\tau^\pm}{2},{\cal U} \right\} \Psi_L\,.
\end{equation}
In the Wigner phase ($\delta_{ph,NG}=0$) one can determine $\eta_{cr}$ by studying the SDE
(\ref{axial_SDE}) for various $\eta$ values.
In NG phase the SDE (\ref{axial_SDE}) at $\eta=\eta_{cr}$ takes the form of a $\tilde\chi$ WTI with NP  breaking term up to $O(b^2)$ that we shall neglect from now on
\begin{equation}\label{WTI_matrix_element}
Z_{\partial\tilde A}\langle 0|\partial_\mu \widetilde  J_\mu^{A^\pm}| M_{PS\pm}\rangle=C_1\Lambda_s \langle 0|{\cal P}^\pm| M_{PS\pm}\rangle.
\end{equation}
Expanding ${\cal U}$ around the vacuum, ${\cal U}=1\!\!1+i\frac{\vec{\tau}\cdot\vec{\varphi}}{v}+O(\frac{\sigma^2}{v^2},\frac{\pi^2}{v^2})$, we get the corresponding expansion for  ${\cal P}^\pm$
\begin{equation}
{\cal P}^\pm=\overline\Psi_L\left\{ {\cal U},\frac{\tau^\pm}{2} \right\} \Psi_R-
h.c.=\overline\Psi_L\left\{1\!\!1+i\frac{\vec{\tau}\cdot\vec{\varphi}}{v}+... ,\frac{\tau^\pm}{2} \right\} \Psi_R-
h.c. =P^{\pm}+...\,.
\end{equation} 
$\chi$ invariance implies that  ${\cal P}^{\pm}$ has the same renormalization constant as $P^\pm=\overline\Psi \gamma_5\tau^{\pm}\Psi$ which we call $Z_P$. Thus a renormalized measure of the effective NP $\widetilde \chi$ breaking is given by 
the dimensionful quantity
\begin{equation}
2m_{AWI}^{ren}\equiv\frac{Z_{\partial\tilde A}}{Z_P}\frac{\langle 0|\partial_\mu \widetilde  J_\mu^{A^\pm}| M_{PS\pm}\rangle}{ \langle 0|{ P}^\pm|M_{PS\pm}\rangle}={C_{1,ren}\Lambda_s} (1+...)\quad\quad 
{C_{1,ren}=\frac{C_1}{Z_P}}\label{m_AWI}
\end{equation}
In spite of its name $m_{AWI}^{ren}$ is not the renormalized counterpart of any parameter in the lattice action. Since at $\eta=\eta_{cr}$ in Wigner phase the current $\tilde J^{A^\pm}_\mu$ is conserved up to $O(b^2)$, $\tilde \chi$-current
algebra implies vanishing anomalous dimension for $\tilde J^{A^\pm}_\mu$ and $\tilde J^{V^3}_\mu$, hence at $\eta=\eta_{cr}$ the NP term on the r.h.s. of the (\ref{WTI_matrix_element}) is RG-invariant.

In principle the study of SDE (\ref{axial_SDE}) would involve evaluation of disconnected diagrams due to isospin changing mediated by the field $\Phi$.
However in the quenched approximation one can prove
 by duplicating the fermion content $\Psi_\ell=(u,d)$ and $\Psi_h=(c,s)$, and considering the SDE involving a generation off diagonal $\tilde J^{A\pm}_\mu$ current,
that eg.~(\ref{axial_SDE}) holds for fermionic disconnected and connected diagrams 
separately. Hence both the determination of $\eta_{cr}$ in Wigner phase and the evaluation of $m_{AWI}$ and related quantities in the NG phase can be carried out in practice without computing fermionic disconnected diagrams.

\section{Renormalisation procedure}
\label{Renormalization}
In a quenched lattice study the renormalization condition on the action parameters can be chosen such that the tuning of $\eta$ to its critical value, the renormalization of  the gauge coupling and the renormalization of the scalar squared mass and quartic
coupling can be carried out separately from each other. For instance:
the relation between $g_0^2 \equiv 6/\beta$ and the lattice spacing is determined by keeping fixed the Sommer length scale $r_0$ in physical units \citep{Guagnelli:1998ud,Necco:2001xg}; the bare scalar mass $m^2_0$ and quartic coupling  $\lambda_0$ are  determined
by keeping fixed (as $b \to 0$ together with $g_0 \to 0$) both $M_\sigma r_0$, where $M_\sigma$
is the mass of the non Goldstone boson scalar particle in the NG-phase, and a suitable non-perturbative definition of the renormalized quartic scalar coupling,  $\lambda_{NP}=M_\sigma^2 /(2v_R^2)$ . In the NG phase the scalar vev, $v_R = Z_\phi^{1/2} \langle \Phi^0\rangle  $, with $Z_\phi$
 the renormalization constant of $\phi$, is thus also fixed in physical units.
The $\phi$-field renormalization constant is computed enforcing 
\begin{equation}
Z_\phi^{1/2}=[M_\sigma \langle\phi^0_{p=0}\phi^0_{p=0}\rangle]-V\langle \Phi^0\rangle ^2]^{-1}\,,
\end{equation}
 the mass $M_\sigma$ is extracted from exponential decay in time correlator $\langle\Phi^0_{x_0}\Phi^0_{0}\rangle$, then $\lambda_R$ is computed.\\
In order to eliminate the dependence from $Z_P$ in the (\ref{m_AWI}) we define the following quantity
\begin{equation}
z_{AWI}^{ren}=2m_{AWI}r_0Z_{\partial\tilde V}G_{PS}^{Wigner}r_0^2
\end{equation}
 where $G_{PS}^{Wigner}\equiv \langle 0 | P^{\pm}|\mbox{PS-meson}\rangle$ is the matrix element of $P^{\pm}$ between vacuum and the pseudo-scalar meson in the Wigner phase of the theory at $\eta=\eta_{cr}$. Note that $z_{AWI}^{ren}=0$ if and only if 
 $m_{AWI}^{ren}=0$.  \\
 The $\chi_L\otimes \chi_R$ symmetry implies that the renormalization constant $Z_{\partial\tilde A}$
 is equal to $Z_{\partial\tilde V}$, the renormalization factor of the vector current (\ref{J_V}). 
 At $\eta=\eta_{cr}$  we have that the vector $\chi$-SDE in the NG phase reads
 \begin{equation}
 Z_{\partial\tilde V}\langle\tilde J^{V,i}_{\mu}(x) \hat{\cal O}(0)\rangle=\langle \Delta\hat{\cal O}\rangle\delta(x)+ C_1\Lambda_S \langle {\cal S}^1(x) \hat{\cal O }(0)\rangle + 2\mu \langle P^1(x) \hat{\cal O}(0)\rangle
 \end{equation}
 where ${\cal S}^1=\overline \Psi_L \left[{\cal U}, \frac{\tau^i}{2}\right]\Psi_R-h.c.$. Taking  $\hat{\cal O}=P^1$ and exploring parity we have
 \begin{equation}
\sum_{\bar x} Z_{\partial\tilde V}\langle\tilde J^{V,i}_{\mu}(x) P^1(0)\rangle=\sum_{\bar x} 2\mu \langle P^1(x) P^1(0)\rangle\,,
 \end{equation}
 since $\sum_{\bar x}\langle {\cal S}^1(x)P^1(0)\rangle=0$.
 The latter equation can be used to determine $Z_{\partial\tilde V}$ which is equal to $Z_{\partial\tilde A}$.
\section{Acknowledgements} 
We acknowledge support from INFN, via the convention INFN-Cineca which made available to us the
CPUtime for carrying out numerical simulations on Galileo, Marconi A1 and Marconi A2 clusters.
Support from the Sino-German CRC110 research network is also gratefully acknowledged.
%We acknowledge support from INFN, via the convention INFN-Cineca 
%which made available to us the CPUtime for carrying out numerical
%simulations on Galilelo, Marconi A1 and Marconi A2 clusters. 
%Support from Centro Fermi, under the contract ... for P.D., and
%from the Sino-German CRC110 research network is also gratefully
%acknowledged.
\clearpage
\bibliography{lattice2017}

%%%%%%%%%%%%%%%%%%%%%%%%%%%%%%%%%%%%%%%%%%%%%%%%%%%%%%%%%%%%%%%%%%%%%%%%%%%%%
\end{document}